\documentclass[fleqn,10pt]{wlscirep}
\usepackage{amsmath}
\usepackage{amssymb}
\usepackage{url}
\usepackage[Large,bf,nooneline]{subfigure} 
\usepackage{multirow}
\usepackage{graphicx}
\usepackage[super,sort&compress]{natbib}
\hypersetup{citecolor=black}

\title{Muon dynamic radiography of density changes induced by hydrothermal activity at the La Soufri\`ere of Guadeloupe volcano.\thanks{Paper submitted to \textsc{Scientific Reports} -- Nature, June 2016}}

\author[1,4,$\dag$]{Kevin Jourde}
\author[2,3,$\ddag$]{Dominique Gibert}
\author[4]{Jacques Marteau}
\author[2]{Jean de Bremond d'Ars}
\author[1,3]{Jean-Christophe Komorowski}

\affil[1]{Institut de Physique du Globe de Paris (CNRS UMR 7154), Sorbonne Paris Cit\'{e}, Paris, France. }
\affil[2]{OSUR - G\'{e}osciences Rennes (CNRS UMR 6118), Universit\'{e} Rennes 1, Rennes, France.}
\affil[3]{National Volcano Observatories Service, Institut de Physique du Globe de Paris (CNRS UMR 7154), Paris, France. }
\affil[4]{Institut de Physique Nucl\'eaire de Lyon, Univ Claude Bernard (UMR 5822 CNRS), Lyon, France.}
\affil[$\dag$]{jourde@ipgp.fr}
\affil[$\ddag$]{gibert@univ-rennes1.fr}

\keywords{Cosmic muons, Tomography, Monitoring, Volcanic hydrothermal system}

\begin{abstract}
Imaging geological structures through cosmic muon radiography is a newly developed technique particularly interesting in volcanology. Here we show that muon radiography may be efficient to detect and characterize mass movements in shallow hydrothermal systems of low-energy active volcanoes like the La Soufri\`ere lava dome. We present an experiment conducted on this volcano during the Summer $2014$ and bring evidence that huge density changes occurred in three domains of the lava dome. Depending on their position and on the medium porosity the volumes of these domains vary from $1 \times 10^6 \; \mathrm{m}^3$ to $7 \times 10^6 \; \mathrm{m}^3$. However, the mass changes remain quite constant, two of them being negative ($\Delta m \approx -0.6 \times 10^9 \; \mathrm{kg}$) and a third one being positive ($\Delta m \approx +2 \times 10^9 \; \mathrm{kg}$). We attribute the negative mass changes to the formation of steam in shallow hydrothermal reservoir previously partly filled with liquid water. This coincides with the apparition of new fumaroles on top of the volcano. The positive mass change is synchronized with the negative mass changes indicating that liquid water probably flowed from the two reservoirs invaded by steam toward the third reservoir.
\end{abstract}
\begin{document}

\flushbottom
\maketitle
\thispagestyle{empty}

\section*{Introduction}

The La Soufri\`ere of Guadeloupe volcano belongs to the Lesser Antilles volcanic arc formed by the subduction of the North American Plate beneath the Caribbean Plate. La Soufri\`ere is the last lava dome in a series of dome extrusions and collapses \cite{komorowski_guadeloupe_2005, boudon_volcano_2007, boudon_new_2008}. During the last 8500 years, about half of the 8 collapses that occurred at La Soufri\`ere of Guadeloupe have generated laterally-directed explosions caused by depressurisation (volcanic blasts) of magma and hydrothermal fluids that spread laterally at high speeds (up to $ 100-235 \; \mathrm{m.s}^{-1}$) over the volcano flanks. These events caused partial collapses and triggered small laterally-directed hydrothermal explosions associated with significant exurgence of hot acid pressurized hydrothermal fluids contained in superficial reservoirs located inside the lava dome \cite{komorowski_guadeloupe_2005, legendre_reconstruction_2007}.

\begin{figure}
\begin{center}
\includegraphics[width=1.0\linewidth]{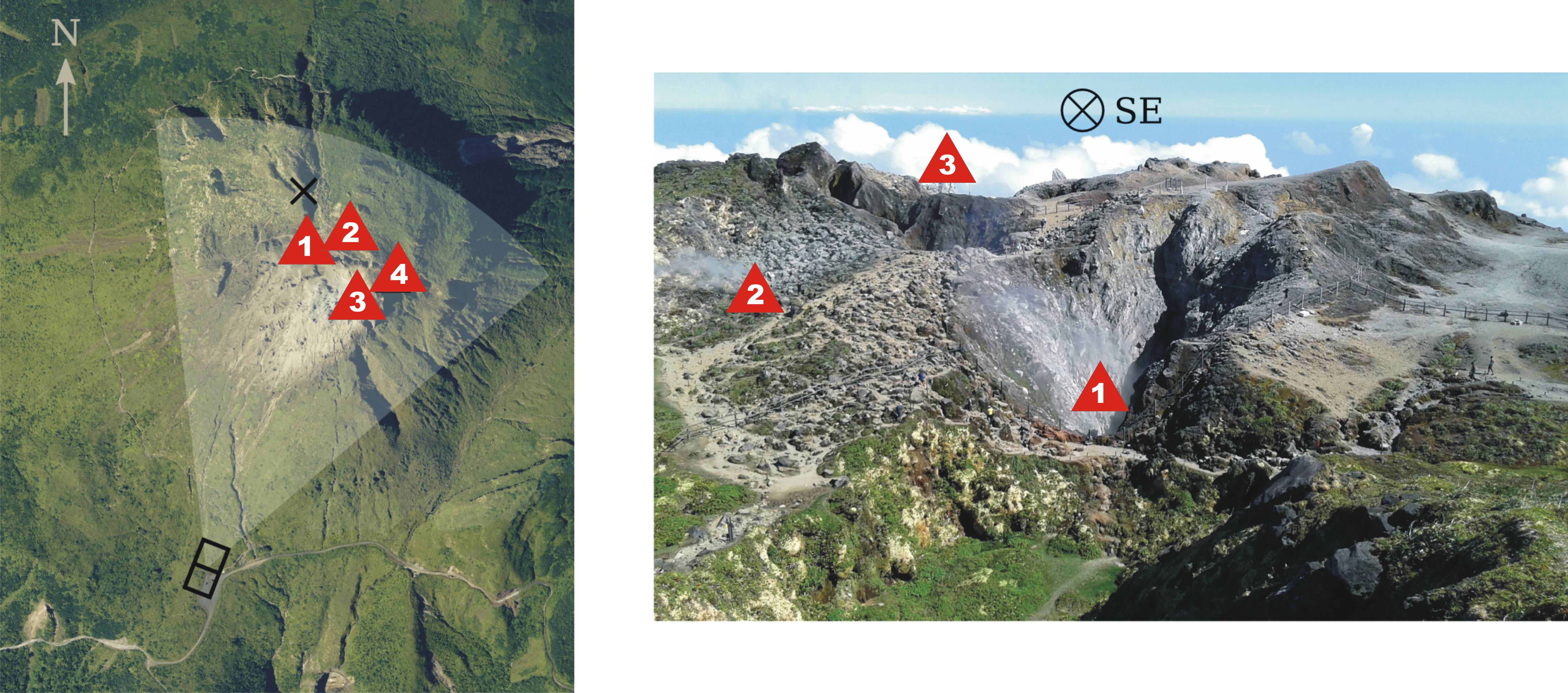}
\end{center}
\caption{\textit{Left~:} Top view of the La Soufri\`ere of Guadeloupe (photography by Institut National de l'Information Géographique et Forestière). The white surface delimits the region scanned by the muon telescope represented by the black boxes. The red triangles refer to the different active zones visible at the dome surface (1~: Tarissan pit, 2~: Napol\'eon-Nord fume, 3~: Crat\`ere Sud, 4~: Gouffre Breislack and Gouffre 56). \textit{Right~:} The La Soufri\`ere's different active vents as seen from the black cross on the left picture.}
\label{figure_1}
\end{figure}

The last eruption occurred in 1976-1977 and is considered as a failed magmatic event where a small andesitic magma volume stopped its ascension about 3 km beneath the surface \cite{feuillard_1975-1977_1983, villemant_evidence_2014}. Following this event, degassing, thermal flux and seismicity progressively decreased down to their lowest levels in 1990. By the end of 1992, a resumption of the fumarolic activity at the summit of the dome, of the shallow seismicity, and of the temperature of thermal springs around the dome was observed. In 1998, the sudden onset of high-flux chlorine degassing from the Crat\`ere Sud vent constituted a conspicuous change in the magmatic-hydrothermal system behaviour. In 1997, a small boiling pond formed at the Crat\`ere Sud with extremely acid fluids (minimal pH equals -0.8) corresponding to the azeotrope point of hydrochloric acid. A similar larger boiling lake was discovered in 2001 in the Tarissan pit. In 2003, the acid pond at the Crat\`ere Sud was replaced by a strongly degassing vent while Tarissan acid pond continues to exist until now. Since then, a significant increase of the dome fumarolic activity was observed\cite{allard_steam_2014,ovsg_bilan_2015, ovsg_bilan_2016}. Indeed a new active region appeared to the North-East of the Tarissan pit during the 2014 Summer \cite{ovsg_bilan_2015}, the North-Napoleon fumarole, and two old pits, the Gouffre Breislack and the Gouffre 56, have seen their activity rising \cite{ovsg_bilan_2016}. All the vents positions are reported on Fig.~\ref{figure_1}.

The spatio-temporal evolution of active areas located onto the lava dome may be caused by the progressive sealing of previously active flow paths. This sealing results from the combined effects of hydrothermal activity and heavy rains $(\approx 6-7 \; \mathrm{m.y}^{-1})$ that favour fluid mineralization by magmatic gas and the formation of clayey material that progressively fills open fractures. Sealing causes fluid confinement and over-pressurization leading to the opening of new flow paths. The observed increased flux of chlorine-rich vents at the summit and the episodic chlorine spikes recorded in the Carbet and Galion hot springs \cite{villemant_evidence_2014} can be due to the sporadic injection of acid chlorine-rich fluids and heat from the magma reservoir or magma intrusions at depth into the hydrothermal reservoirs \cite{villemant_evidence_2014}.

Purely hydrothermal blasts are as mobile as their magmatic counterparts. Small-scale laterally-directed explosions could have caused many fatalities at Tongariro (New-Zealand) in 2012 \cite{hurst_precursory_2014,jolly_overview_2014, lube_dynamics_2014} and caused 63 fatalities at Ontake volcano (Japan) in 2014 \cite{cyranoski_why_2014,kato_preparatory_2015,maeda_source_2015,sano_ten-year_2015}. In both cases, no clear warning signals were reported. Both the detection of early warning signals, the quantification of both the hydrothermal volumes involved and the amount of energy available, constitute new challenges to modern volcanology. As quoted by several experts who commented the Ontake eruption, theses challenges deserve the development of new techniques providing new data type and of new concepts in data analysis and information processing.

The aim of this paper is to present the recently developed density muon radiography technique used to monitor density changes in the shallow hydrothermal system inside the lava dome of the La Soufri\`ere of Guadeloupe. As will be shown in the remaining of this paper, it allows to clearly detect density changes related to modifications of the vent activity visible at the volcano's summit. Quantitative attributes like time-constants, concerned volumes and estimates of the amount of thermal energy available may be derived from muon radiography data.

The paper is organized as follows. First, in order to make the paper self-consistent, we recall the main principles of cosmic muon radiography. Then we present the experiment performed on the La Soufri\`ere of Guadeloupe during Summer $2014$, while the North-Napoleon fumarole appeared. In a second stage, we present the data and, after eliminating the possibility of artefacts caused by atmospheric perturbations, we provide evidence of large mass movements within the lava dome. Finally, we briefly discuss the consequences for the dynamics of the shallow hydrothermal system.

\section*{The muon radiography experiment}

Muon radiography aims at recovering the density distribution, $\rho$, inside the volcano by measuring its screening effect on the cosmic muons flux \cite{george_cosmic_1955, alvarez_search_1970, nagamine_introductory_2003, tanaka_subsurface_2013}. The material property inferred with muon radiography is the opacity, $\varrho$ $[\mathrm{g.cm}^{-2}]$, which measures the amount of matter encountered by the muons along their travel path, $L$, across the rock mass to image,
\begin{equation}
\varrho = \int_L \rho(l) \times \mathrm{d}l.
\label{opacity1}
\end{equation}

Muons lose energy through matter by ionisation processes at a typical rate of $2.5 \; \mathrm{MeV}$ per opacity increment of $1 \; \mathrm{g.cm}^{-2}$. They travel along straight trajectories across low-density materials, including rocks, and scattering is significant only for low-energy muons ($E_c \le 1 \; \mathrm{GeV}$). Muon radiography of kilometer-size objects like volcanoes involves the hard muonic component with energy above several hundredths of $\mathrm{GeV}$. In such cases, the muons incident flux is nearly stationary, azimuthally isotropic and mainly depend on the zenith angle \cite{gaisser_cosmic_1990, lesparre_geophysical_2010}. Simple flux models can be used to determine the target screening effects\cite{tang_muon_2006, lesparre_geophysical_2010}. These properties have been used for 10 years to image volcanoes internal structures \cite{tanaka_imaging_2007, lesparre_density_2012}, and more recently to monitor mass variations linked to their activity \cite{gibert_density_2013, tanaka_radiographic_2014}.

Our cosmic muon telescopes \cite{lesparre_design_2012} are equipped with 3 plastic scintillator matrices counting $16 \times 16$ pixels of $5 \times 5 \; \textrm{cm}^{2}$. We set the distance between the front and rear matrices to $140 \; \textrm{cm}$ in order to span the entire lava dome from a single point of view (Fig.~\ref{figure_1}). The combination of pixels defines a total of $31 \times 31$ lines of sight with a spatial resolution of about $25 \; \textrm{m}$ at the lava dome center. A $5 \; \textrm{cm-thick}$ lead shielding is placed just behind the center matrix. The matrices geometrical arrangement fixes the telescope acceptance function $\mathcal{T}_{m} \; [\mathrm{cm}^2.\mathrm{sr}]$ which controls the flux captured by the instrument on each line of sight $\mathbf{r}_{m}$. As shown below, the acceptance has a direct influence on the number of muons counted in a given period of time and, consequently, controls the statistical uncertainty of eventual flux variations caused by density changes in the volcano \cite{lesparre_geophysical_2010, jourde_monitoring_2016}.

The present dataset runs from July $7^\mathrm{th}$ to August $5^\mathrm{th}$ and from August $15^\mathrm{th}$ to October $10^\mathrm{th}$ $2014$. The telescope was placed South-South-West at the lava dome basis (Fig.~\ref{figure_1}), the axial line of sight was oriented $\beta_{0} = 25^\circ$ Eastward and the zenith angle $\gamma_{0} = 75^\circ$ (the telescope is horizontal when $\gamma_{0} = 90^\circ$).

The raw data processing has been standardized. Data reduction and filtering include~: events time coincidence (on the $3$ matrices), particle time-of-flight, alignment of the fired pixels, number of pixels activated in the rear matrix. This number is larger when particles other than muons (electrons and gamma photons) start showering in the lead shielding. The resulting data set is a sequence $\mathcal{S} = \{ e_k, \; k = 1,\cdots,K \}$ of events $e_k$ attributed to cosmic muons arriving in the telescope front. Each event has a time-stamp $t_k$ and is assigned to one particular line of sight $\mathbf{r}_{m(k)}$.

Once obtained, the sequence $\mathcal{S}$ is used to compute the muons number $N_{m}(t,\Delta t)$ for each line of sight $\mathbf{r}_{m}$ during a time period $\Delta t$. It must be corrected from the acceptance function $\mathcal{T}_{m}$ to suppress the instrument efficiency.

An additional processing stage specific to the present study (see Methods section below) consists in merging several adjacent lines of sight in order to improve the signal-to-noise ratio. For lines belonging to a subset $\mathcal{E}$,
\begin{equation}
N_{\mathcal{E}}(t,\Delta t) = \sum_{m \in \mathcal{E}} N_m(t,\Delta t) = \sum_{m \in \mathcal{E}} \mathcal{T}_m \times \int_{t - \Delta t / 2}^{t+\Delta t / 2} \partial \phi_m(\zeta) \mathrm{d}\zeta,
\label{N_particles_detected}
\end{equation}
where $\mathcal{T}_m$ and $\partial \phi_m$ are respectively the acceptance function and the flux in the line of sight $\mathbf{r}_{m}$. In the following we use $\phi_{\mathcal{E}} (t,\Delta_t) = N_{\mathcal{E}}(t,\Delta t) / \Delta t$ the particle flux at $t$, averaged on a $\Delta t$ time window. As explained in the Methods section, the lines of sight belonging to a given subset $\mathcal{E}$ must share a similar time-variation curve of the muons flux. Such a merging increases the effective acceptance and improves the time resolution. The counterpart is a decrease in the angular resolution induced by the merging of the small solid angles spanned by the lines of sight.

The telescope mechanical stability is an important issue for monitoring experiments. Changes in the telescope orientation may occur because of the strong mechanical constraints the instrument has to support, especially for measurements under open sky where about $300 ~ \mathrm{kg}$ of steel shielding are necessary as explained above. The ground below the telescope may also slightly move as the Soufri\`ere of Guadeloupe is subject to heavy rains all the year long. Periodic check of both the orientation and inclination of the telescope frame did not reveal variations during the experiment. The overall telescope detection readout efficiency is monitored permanently on the data themselves (using the responses of open-sky oriented lines) and thanks to dedicated open-sky calibration runs. No significant changes have ever been observed. This is also confirmed by the Principal Component Analysis (PCA, see Methods section below) used to merge the lines of sight into coherent domains, since it is able to identify and remove global shifts in all lines of sight. Such an effect was not detected during the processing, we thus conclude that no instrumental bias alter the data.

\section*{Evidence of opacity changes inside the La Soufri\`ere lava dome}

\begin{figure}
\begin{center}
\includegraphics[width=0.7\linewidth]{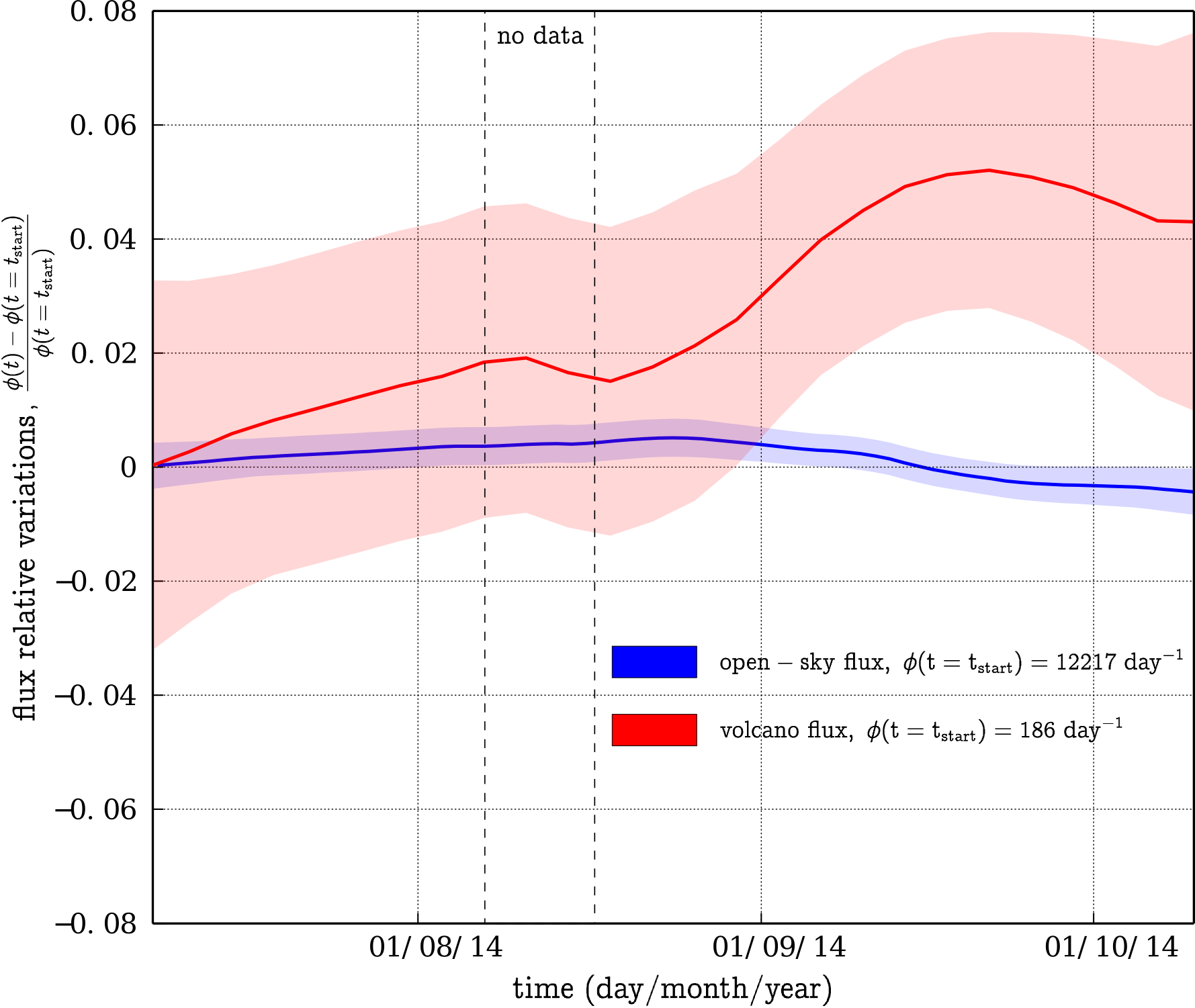}
\end{center}
\caption{Relative muon flux measured during Summer 2014. In blue~: open-sky axes. In red~: axes pointg through the La Soufri\`ere of Guadeloupe volcanic dome. The transparent surfaces associated to each curve delimit the $95\%$ confidence interval. The fluxes are computed using a 30 days large Hamming moving window. The vertical black dotted lines delimit the small period during which the muon telescope was not operational.}
\label{figure_2}
\end{figure} 

The red curve in Fig.~\ref{figure_2} shows the muon flux global relative variations through the La Soufri\`ere lava dome. This curve has been obtained by applying eq. \ref{N_particles_detected} to the subset $\mathcal{E}_\mathrm{dome}$ of all lines of sight that pass through the volcano. The blue curve of Fig.~\ref{figure_2} corresponds to the subset $\mathcal{E}_\mathrm{sky}$ of the lines of sight directed toward the open-sky above the volcano. The transparent surfaces represent the $95\%$ confidence interval. The flux data were smoothed over a $30$ days large Hamming window. This smoothing interpolates the data gap delimited by the two vertical black dotted lines. Fig.~\ref{figure_2} red curve shows a conspicuous increase of about $5\%$ of the muon flux during the whole observation period. Such an increase is not visible in the open-sky flux which instead slightly decreases by about $1\%$ (Fig.~\ref{figure_2} blue curve). This indicates that the muons flux increase across the lava dome is likely to be primarily caused by a decrease of the volcano bulk opacity.

\begin{figure}
\begin{center}
\includegraphics[width=0.8\linewidth]{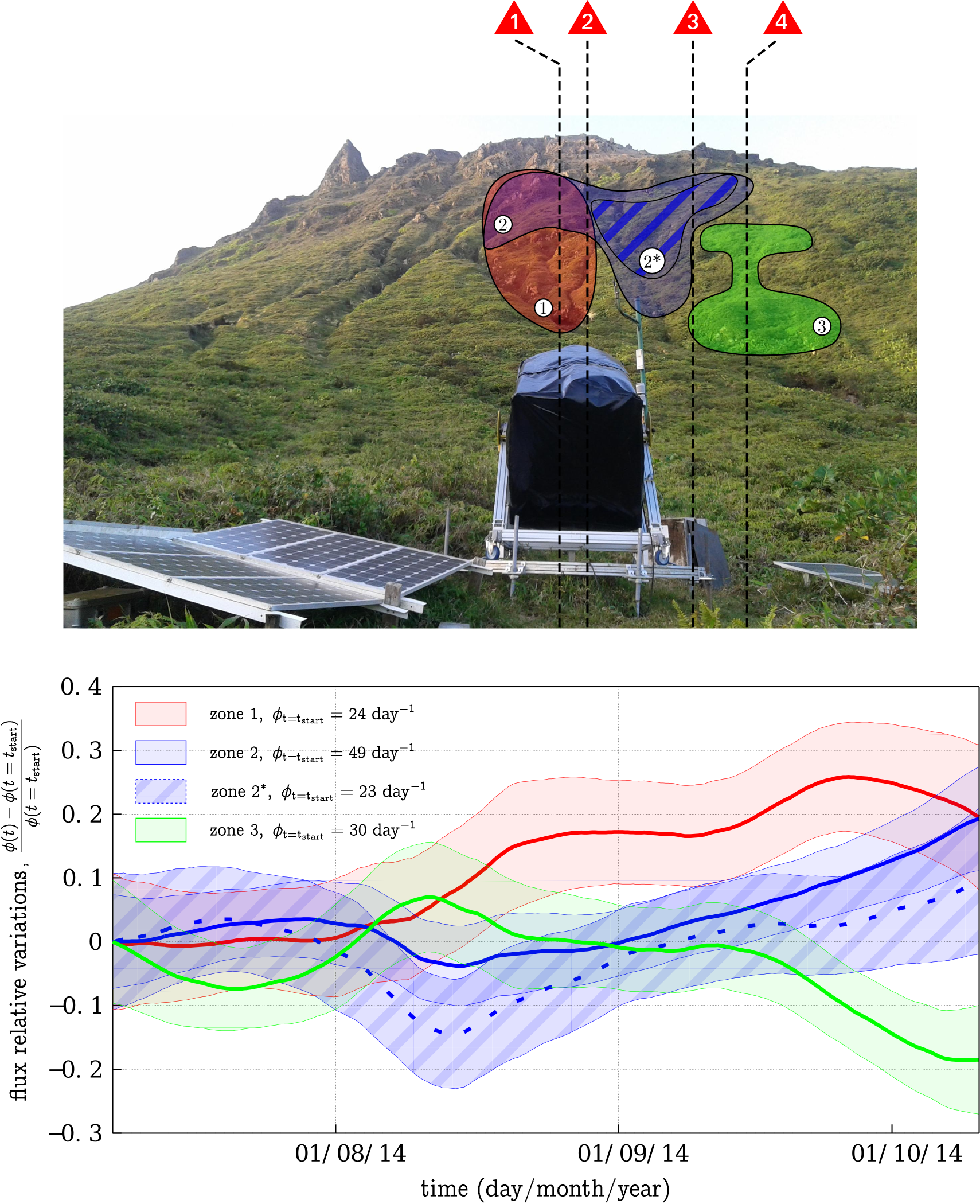}
\end{center}
\caption{\textit{Top~:} The La Soufri\`ere of Guadeloupe as seen from the muon telescope. The coloured regions refer to the different regions from which we could extract coherent temporal signals. The red triangles refer to the different active zones visible at the surface of the dome azimuths (Fig.~\ref{figure_1}). \textit{Bottom~:} The relative muon flux variations associated to the top picture regions during Summer 2014. The transparent surfaces associated to each curve delimit their respective uncertainty with a $95\%$ confidence interval. The fluxes are computed using a $30$ days large Hamming moving window.}
\label{figure_3}
\end{figure} 

In order to identify the locations in the lava dome where the muon flux variations are the most important, we performed a PCA (see Methods section) on all $N_m(t)$ series. The analysis identifies 3 domains where the muons flux time-variations are similar. Each domain lines of sight were merged to compute $N_{\mathcal{E}}, \; \mathcal{E} \in \{1, 2, 2^*, 3\}$. $2^*$ is a volume $2$ sub-volume that is not aligned with volume $1$ and that consequently only shows the volume $2$ specific variations. Fig.~\ref{figure_3} shows the 3 identified domains and their associated muons flux time series. We observe muons flux relative variations as large as $20\%$. We exclude from the PCA instrumental or atmospheric effects which would have impacted all the telescope observation axes. The unambiguous spatial localization and the diversity of the different temporal trends appear to be the strongest argument in favour of explaining the variations by the volcanic activity itself. Statistical considerations concerning time-resolution and potential conclusions on the lava dome hydrothermal activity are briefly discussed in the next two sections. 

\subsection*{Atmospheric effects}

The cosmic muons are produced in the so-called extended atmospheric showers (EAS) which initiate in the upper atmosphere, at an altitude $z_\mu \approx 15-20 \; \mathrm{km}$, when primary cosmic ray particles (mainly protons) collide with oxygen and nitrogen atoms. When the thermal state of the atmosphere changes, both the air density and $z_\mu$ are slightly modified, producing small variations in the muon flux $\phi$ with respect to its average $\bar{\phi}$. For the present analysis, we use the simplest linear exmpirical relationships \cite{blackett_instability_1938,duperier_new_1944,jourde_monitoring_2016} like,
\begin{equation}
\frac{\phi - \overline{\phi}}{\overline{\phi}} = \beta_p^* \times(p-\overline{p}) + \alpha_T \times \frac{ (T_\mathrm{eff}-\overline{T}_\mathrm{eff}) } { \overline{T}_\mathrm{eff}},
\label{atm_effects_eq_2}
\end{equation}
where $p \; [\mathrm{hPa}]$ is the ground pressure, $\overline{p}$ the local average ground pressure, $\alpha_T = \beta_T^* \times \overline{T}_\mathrm{eff}$ the modulation coefficient and $T_\mathrm{eff}$ the effective temperature. This temperature is an approximation of $T_\mu$,  the atmosphere temperature at the high energy muons production altitude $z_\mu$, not to be misinterpreted with the ground temperature. It corresponds to an estimate of the atmosphere temperature where high-energy muons are produced, at the very beginning of the EAS development. 

The coefficients $\beta_p^*$ and $\beta_T^*$ are computed through linear regressions. They mainly depend on the measurement site altitude and latitude \cite{motoki_precise_2003}, and on the cut-off energy $E_{c} (\varrho)$ corresponding to the screening caused by the amount $\varrho$ of matter facing the telescope \cite{barrett_interpretation_1952}, a primary importance in applications aiming at imaging large geological structures.

The barometric coefficient $\beta_p^*$ is negative since an increase in $p$ induces an increase in the air column opacity. It makes the atmosphere harder to go through for the muons and reduces their total flux. However it mainly affects the soft muons with an energy less than a few $\mathrm{GeV}$.
$\beta_p^*$ is a decreasing function of $E_c$. This is due to the fact that the differential energy spectrum of cosmic muons is a sharply decreasing function. Consequently, as $E_{c} (\varrho)$ increases, the number of low-energy muons stopped by an increase of atmosphere opacity is smaller. In the SHADOW experiment \cite{jourde_monitoring_2016} where $E_c \approx 1~\mathrm{GeV}$, we got $\beta_p^* = 0.0013 ~ (0.0002) \; \mathrm{hPa}^{-1}$. For the present experiment, the opacity ranges from $200$ to $2000 \; \mathrm{mwe}$ and $50 < E_{c} < 675 \; \mathrm{GeV}$. Numerous studies have measured or computed $\beta_p^*$ for various cut-off energies \cite{sagisaka_atmospheric_1986,motoki_precise_2003}, and in our case we have $\beta_p^* (E_c \gtrsim 50 \; \mathrm{GeV})  \lesssim 0.0001$. At the summit of the La Soufri\`ere, $\overline{p} = 858 \; \mathrm{hPa}$ and pressure variations $p - \overline{p} < 5 \; \mathrm{hPa}$. We extract that corresponding muons flux relative variations are less than $0.1\%$, much less than the muon flux variations reported in figures~\ref{figure_2} and~\ref{figure_3}.

The second right-hand term of eq. \ref{atm_effects_eq_2} represents the temperature effect, which has been diversely parametrized in the literature \cite{duperier_new_1944,barrett_interpretation_1952,barrett_atmospheric_1954}. We adopt here the parametrization of Barrett \textit{et. al} \cite{barrett_interpretation_1952,barrett_atmospheric_1954} who define $T_\mathrm{eff}$ as the atmosphere temperature vertical average weighted by the cosmic pions disintegration probability. From the pions interaction probability with the atmosphere molecules (increasing with density) and their decay probability into muons (also affected by the local density), one expects $\beta_T^* ~ \mathrm{[K^{-1}]}$ negative for low energy particles (roughly \cite{adamson_observation_2010} for $E_c$ below $10~\mathrm{GeV}$), and positive for high energy particles.

\begin{figure}
\begin{center}
\includegraphics[width=0.7\linewidth]{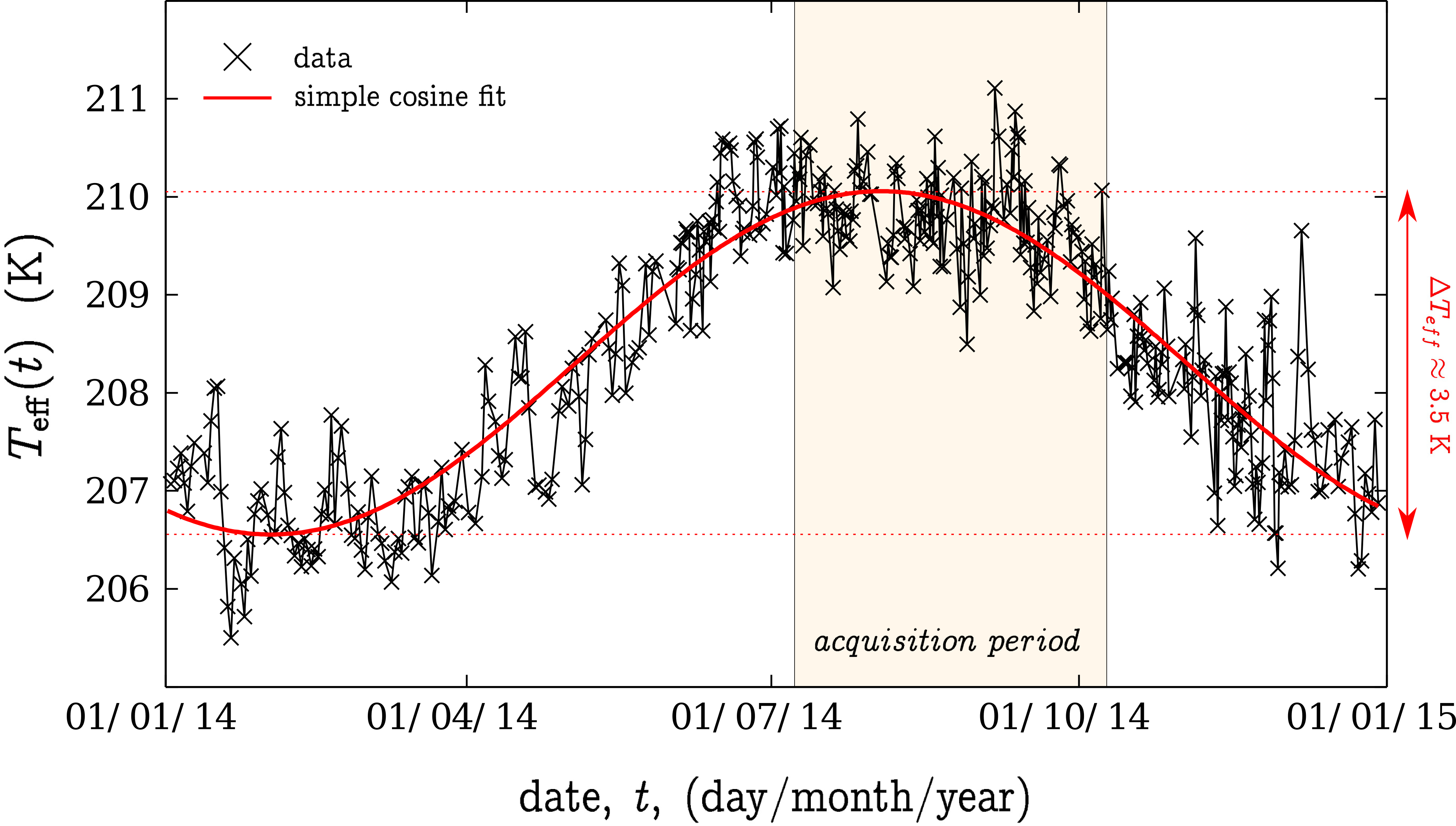}
\end{center}
\caption{Effective temperature $T_\mathrm{eff}$ estimated from Le Raizet daily launched atmospheric sounding balloons (black crosses). The red curve is a simple fit using a cosine function. The data was extracted from the University of Wyoming website (http://weather.uwyo.edu/upperair/sounding.html).}
\label{figure_4}
\end{figure} 

For the volcano experiment discussed here, we expect a negative $\alpha_T$ for the open-sky lines of sight, as most of the detected muons have a low energy and a positive $\alpha_T$ is expected for the lines of sight crossing the volcano. Fig.~\ref{figure_4} shows the $T_\mathrm{eff}$ time series for the whole year 2014 in Guadeloupe using meteorological sounding balloons data. As can be observed, $T_\mathrm{eff}$ is well correlated with the seasons and it variations amplitude is about $\Delta T_\mathrm{eff} \approx 3.5~\mathrm{K}$. This value is very small because the Guadeloupe archipelago is close to the equator ($\lambda = 16.02^{\circ} \; \mathrm{North}$) where seasonal effects are almost absent. It should be compared, for example, to the $T_\mathrm{eff}$ annual variation $\Delta T_\mathrm{eff} \approx 42 ~\mathrm{K}$ observed at the ICECUBE detector in Antarctica ($\lambda = 89.59^{\circ} \; \mathrm{South}$). For the ICECUBE experiment \cite{tilav_atmospheric_2010}, $\alpha_T \approx - 0.36$ for the open-sky muon flux ($E_{c} \approx 1 \; \mathrm{GeV}$), and $\alpha_T \approx + 0.90$ for high-energy muon flux ($E_{c} \approx 400 \; \mathrm{GeV}$).

The present dataset being quite short, the sampling of $T_\mathrm{eff}$ annual variation is affected by large uncertainty~: $\alpha_T = + 1.05 ~ (1.45)$ ($95\%$ confidence interval) for the open-sky flux (Fig.~\ref{figure_2} blue curve). Using the $\alpha_T$ values computed as a function of $E_c$ through numerical models and measured with various underground experiments \cite{tilav_atmospheric_2010, ambrosio_seasonal_1997, adamson_observation_2010} (e.g. Fig.~7 from Adamson \textit{et. al} \cite{adamson_observation_2010}), we obtain  $0.2 < \alpha_T < 0.8$ for the lava dome opacity range $200 ~ \mathrm{mwe} < \varrho < 2000 ~ \mathrm{mwe}$. With these values, the expected relative muon flux variations due to temperature effects should not exceed $0.3\%$ during the acquisition period, still more than one order of magnitude below the detected fluctuations (Fig.~\ref{figure_2} red curve and Fig.~\ref{figure_3}).

From the results discussed in this section, we may safely conclude that neither the pressure effects nor the temperature effects may explain a significant part of the muon flux variations observed through the lava dome.

\subsection*{Statistical time resolution}

The temporal fluctuations one can extract from the muons flux signal are intrinsically limited by the statistical noise. We previously demonstrated \cite{jourde_monitoring_2016} that the minimum acquisition time $\Delta t_\mathrm{min}$ necessary to extract a relative variation $\varepsilon$ from the average detected flux $\phi_\mathcal{E}$ reads,
\begin{equation}
\Delta_t > \Delta t_\mathrm{min} = \frac{\tilde{\alpha}^2 \times (1 - \varepsilon^2 / 4)}{\varepsilon^2 \times \phi_\mathcal{E}},
\label{stats_eq_1}
\end{equation}
with $\alpha = \mathrm{erf} ( \tilde{\alpha} )$ the confidence interval chosen to validate the statistical hypothesis.

A single observation axis that crosses the la Soufri\`ere of Guadeloupe in this experiment detects between 1 and 5 particles per day. Then according to eq.~\ref{stats_eq_1} it can at best extract during our $95$ days experiment a respectively $40~\%$ and $20~\%$ relative flux variation (with a $2 \sigma$ precision). We could not detect such strong fluctuations in la Soufri\`ere of Guadeloupe on this time-scale on a single axis. The solution is then to sum the signals from the different observation axes (see Methods section). Doing so we increase $\phi_\mathcal{E}$, and improve both the time and amplitude resolutions of the fluctuations (respectively $\varepsilon$ and $\Delta t_\mathrm{min}$). As a counterpart, we deteriorate their spatial localization. 

As presented on Fig.~\ref{figure_3}, we are finally able to extract $\varepsilon \approx 10~\%$ relative fluctuations on a $30$ days time-scale in regions regrouping from 21 (with zone $\mathcal{E}_1$) to 42 axes (with zone $\mathcal{E}_2$). Taking an average flux of 3 particles per day per observation axis, we respectively get with eq.~\ref{stats_eq_1} minimum acquisition times ranging from $13$ to $24$ days. These numbers are coherently below the $30$ days Hamming window mentioned previously.

\section*{Discussion and conclusions}

\begin{figure}
\begin{center}
\includegraphics[width=1.0\linewidth]{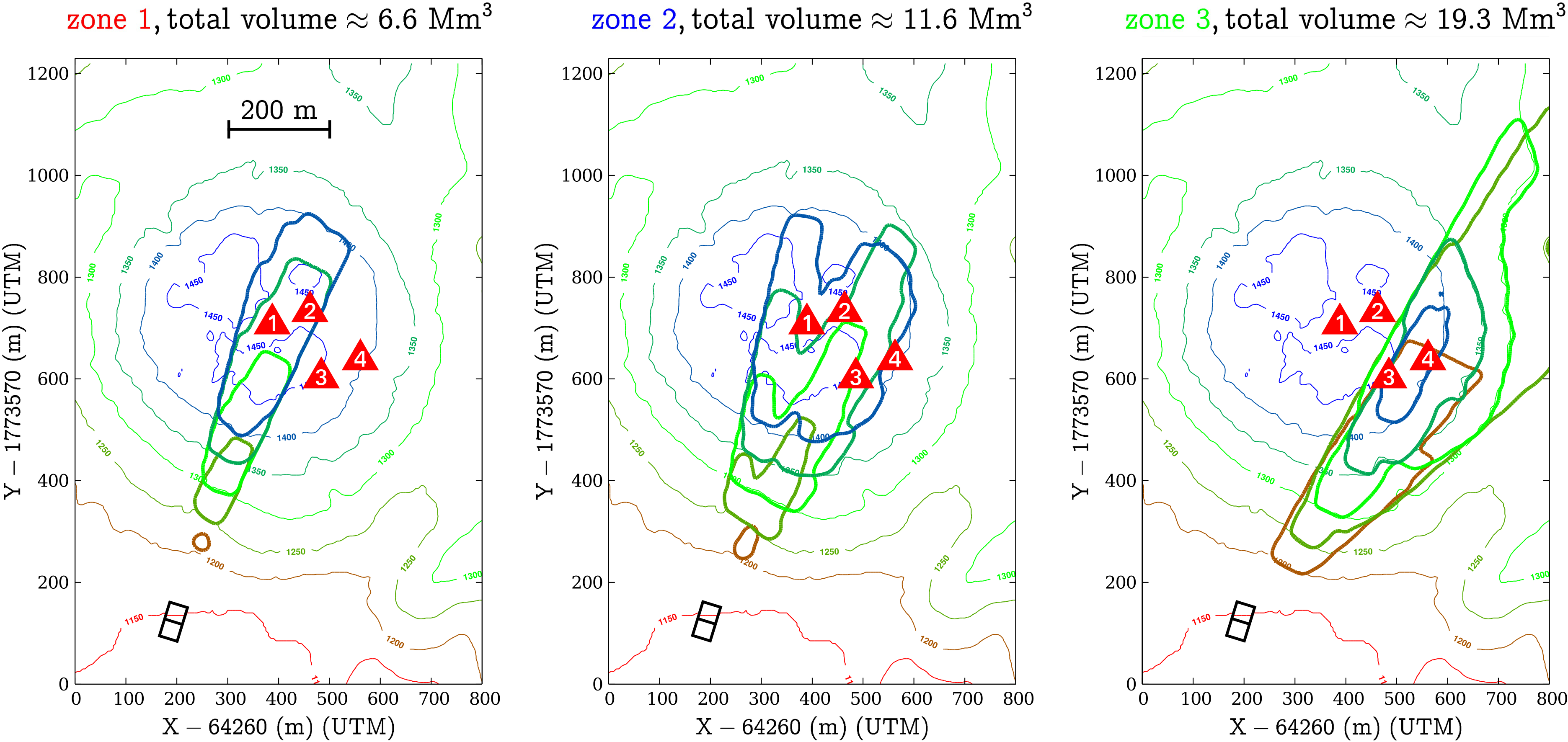}
\end{center}
\caption{Plan view of the volume scanned by the muon telescope for the three domains $\mathcal{E}_1$, $\mathcal{E}_2$ and $\mathcal{E}_3$ represented on Fig.~\ref{figure_3}. The joined squares located SouthWest represent the telescope, and the red triangles refer to the active areas visible on the lava dome summit plateau (Fig.~\ref{figure_1}).}
\label{figure_5}
\end{figure} 

\begin{figure}
\begin{center}
\includegraphics[width=0.9\linewidth]{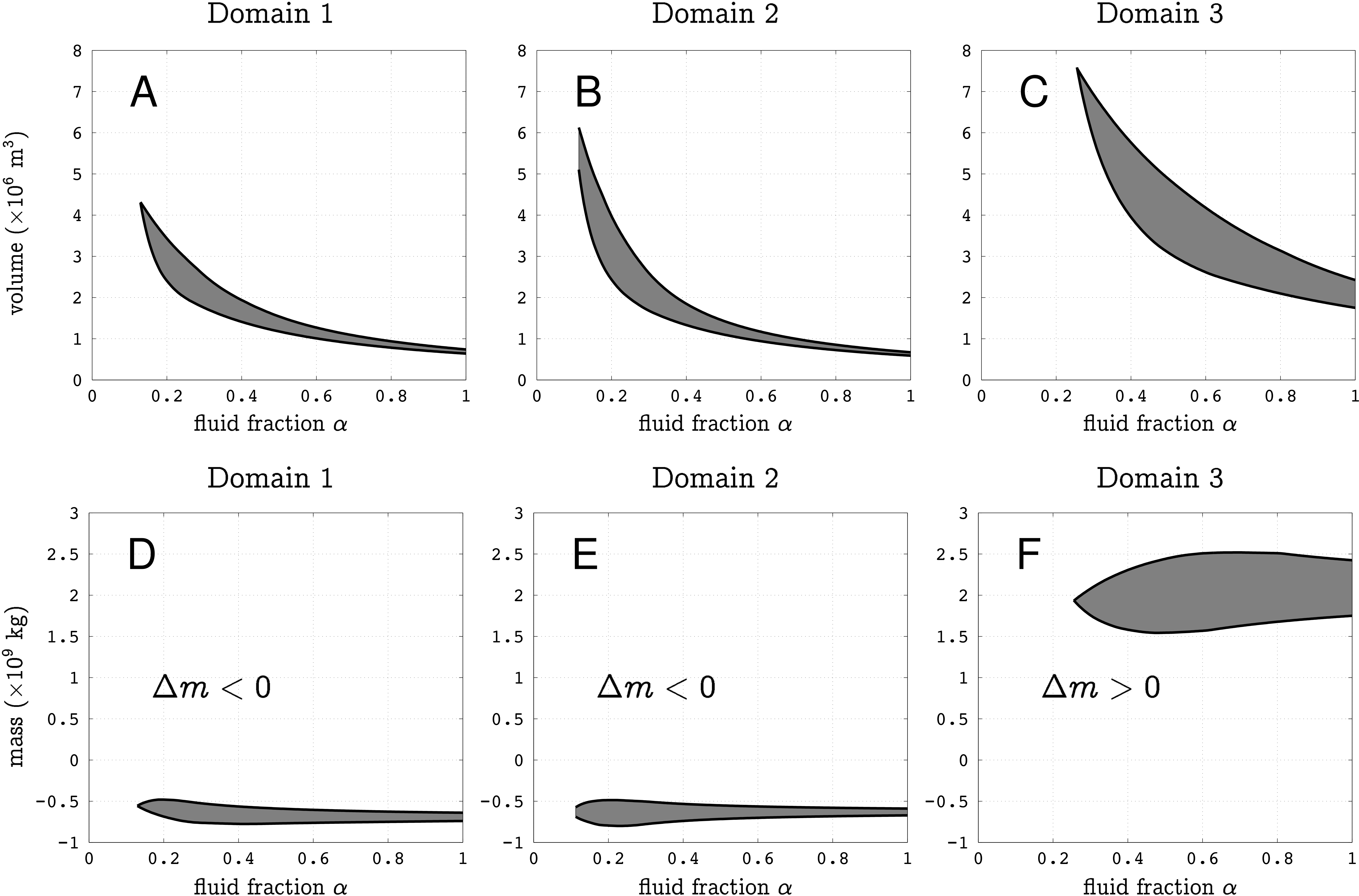}
\end{center}
\caption{\textit{Top~:} Volume ranges $V_\mathcal{E}$ for the domains $\mathcal{E}_1$ (A), $\mathcal{E}_2$ (B) and $\mathcal{E}_3$ (C) as a function of the fluid fraction $\alpha$. \textit{Bottom~:} Ranges of mass change $\Delta m_\mathcal{E}$ for the domains $\mathcal{E}_1$ (D), $\mathcal{E}_2$ (E) and $\mathcal{E}_3$ (F) as a function of the fluid fraction $\alpha$. }
\label{figure_6}
\end{figure} 

\begin{figure}
\begin{center}
\includegraphics[width=0.80\linewidth]{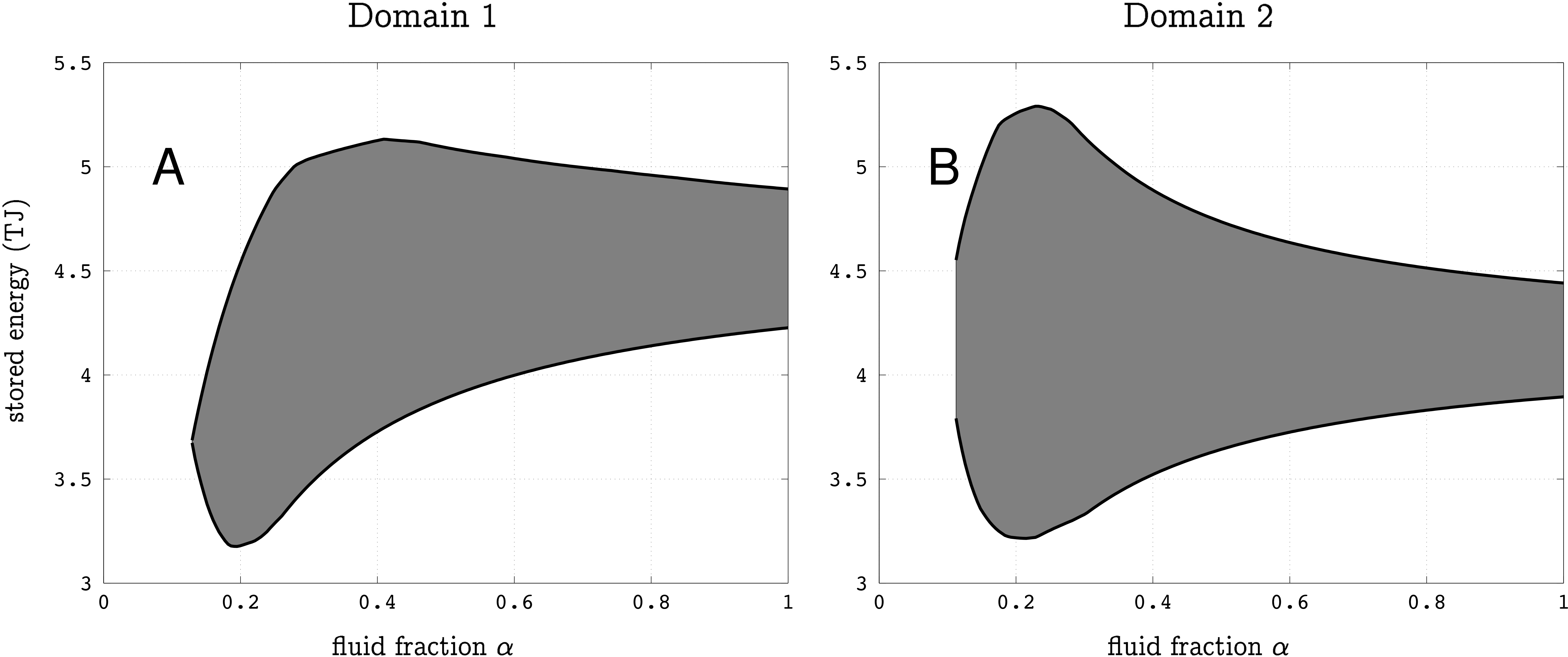}
\end{center}
\caption{Energy brought by steam in domains $\mathcal{E}_1$ (A) and $\mathcal{E}_2$ (B) as a function of the fluid fraction $\alpha$. Values computed for an absolute pressure $p = 6 \; \mathrm{bars}$.}
\label{figure_7}
\end{figure} 

The data presented and discussed in the preceding sections show that huge opacity variations occurred in the La Soufri\`ere lava dome during Summer 2014. Using spatial PCA, theses variations revealed that the observed opacity changes are organized as separate regions located in the volcano southern half underneath the active fumarolic areas visible on the lava dome summit (Fig.~\ref{figure_3}). Both domains $\mathcal{E}_1$ and $\mathcal{E}_2$ show a conspicuous increase of their muon flux roughly starting August $1^\mathrm{st}$ for $\mathcal{E}_1$ and August $10^\mathrm{th}$ for $\mathcal{E}_2$. In the same period, a decrease of the muon flux through domain $\mathcal{E}_3$ starts August $8^\mathrm{th}$. The only plausible explanation is that these rather fast mass changes originate from fluid movements inside the dome \cite{rosas-carbajal_volcano_2016}.

The muon flux variations, i.e. opacity fluctuations, in each domain may be used to estimate their associated mass changes. However, since the lines of sight are small conical volumes with their apex located on the telescope, the mass changes depend on the location of the concerned volume along the line of sight. We draw the entire volumes intercepted by the telescope for the 3 considered lines of sight subsets on Fig.~\ref{figure_5}.  Let us assume that an opacity change $\Delta \varrho_m$ is observed for a given line of sight $\mathbf{r}_m$ and that a density variation $\delta \rho$ occurs somewhere in a volume $V_m$ along this line. The segment length $L_m$ of $V_m \cap \mathbf{r}_m$ reads (see eq. \ref{opacity1}),
\begin{equation}
L_m = \Delta \varrho_m / \delta \rho.
\label{InterVolumeSight}
\end{equation}
If $V_m$ is located at a distance $L_0$ from the telescope,
\begin{equation}
V_m = \Omega_m \int_{L_0-L_m/2}^{L_0+L_m/2} l^2 \mathrm{d}l,
\label{VolumeSight}
\end{equation}
where $\Omega_m$ is the small solid angle spanned by the line of sight $\mathbf{r}_m$. The total volume $V_\mathcal{E}$ is obtained by summing the volumes of all lines of sight belonging to a given domain $\mathcal{E}$,
\begin{equation}
V_\mathcal{E} = \sum_{\mathbf{r}_m \in \mathcal{E} } V_m,
\label{VolumeE}
\end{equation}
and the mass change associated with $V_\mathcal{E}$ is given by,
\begin{equation}
\Delta m_\mathcal{E} = V_\mathcal{E} \times \delta \rho.
\label{MassVolume}
\end{equation}

To compute the volumes $V_\mathcal{E}$ and their corresponding mass changes $\Delta m_\mathcal{E}$, we need to know both $\delta \rho$ and $L_0$. The density change is assumed to be caused by fluid movements where liquid is replaced by gas (air or steam) or inversely. In such a case, $\delta \rho = \alpha \times 1000 \; \mathrm{kg.m}^{-3}$, where $\alpha$ is the volume fraction occupied by the fluids in the rock matrix. The choice of $L_0$ may be guided by geological information about the locations where fluid movements are likely to occur inside the lava dome. In the present study, we consider that fluid movements occur under the active areas that occupy the Southern-East quarter of the summit plateau \cite{rosas-carbajal_volcano_2016} (see the red triangles on Fig.~\ref{figure_1}). The volumes and mass changes corresponding to $\mathcal{E}_1$, $\mathcal{E}_2$ and $\mathcal{E}_3$ are shown in Fig.~\ref{figure_6}.

It may be expected that the positive mass variation of $\mathcal{E}_3$ (Fig.~\ref{figure_6}F) is due to the filling of a perched aquifer with rain water. Taking a surface $S = 1.5 \times 10^5 \; \mathrm{m}^2$ for the summit plateau of the lava dome and a total amount of rain of $2 \; \mathrm{m}$ during August and September, we obtain an upper bound of $S = 3 \times 10^8 \; \mathrm{kg}$ of water input on top the lava dome. However, most of this water is rapidly evacuated through run-off at the surface and it is very unlikely that filling $\mathcal{E}_3$ with rain water is the cause of the positive mass variation $\Delta m_3$. A more reasonable hypothesis is that this mass movement is due to fluid flow from $\mathcal{E}_1$ and $\mathcal{E}_2$ into $\mathcal{E}_3$. This is sustained by the fact that the positive mass variation $\Delta m_3$ is anti-correlated with the negative time-variations of mass associated with $\mathcal{E}_1$ and $\mathcal{E}_2$ (Fig.~\ref{figure_3}). Actually, the mass-variation ranges (Fig.~\ref{figure_6}) allow to have an almost vanishing mass budget, with the mass increase in $\mathcal{E}_3$ compensating the decreases in $\mathcal{E}_1$ and $\mathcal{E}_2$. In such a case, we expect that high-pressure steam formed in $\mathcal{E}_1$ and $\mathcal{E}_2$ and pushed liquid into $\mathcal{E}_3$. Considering that the steam pressure must be sufficient to push liquid and assuming a maximum reservoir depth of $50 \; \mathrm{m}$, we take an absolute pressure $p = 6 \; \mathrm{bars}$. This gives a thermal energy storage of about $4.5 \times 10^{12} \; \mathrm{J}$ in each reservoir $\mathcal{E}_1$ and $\mathcal{E}_2$ (Fig.~\ref{figure_7}).

To conclude, PCA appears to be a fitted solution to regroup the telescope observation axes. We could extract 3 clear domains associated with various temporal signals that are probably linked to the different physical processes occurring into the volcano. PCA being a linear decomposition, it permits to consider an alignment of different zones on a single observation axis which is an intrinsic problem to tomography imaging from a single point of view. It is the case here with the zones $\mathcal{E}_1$ and $\mathcal{E}_2$ (Fig.~\ref{figure_3} and Fig.~\ref{figure_8}). The different domains geometrical shapes, as well as their alignment with the surface active vents are probably the best argument in favour of the detection of a volcanic signal because no prior information was injected into the analysis. $\mathcal{E}_2$ even shows a clear connection between two active regions.

However let us mention three factors that limit this approach. First our analysis assumes the temporal trends occur in fixed areas while we can expect the regions to change size, to merge in between each others, or to divide into smaller independent ones. Second the PCA algorithms searches the simplest possible solution. Many other combinations can be imagined, and for example two regions that completely superimpose onto the other as seen from the telescope cannot be distinguished (Fig.~\ref{figure_3}). And third PCA zoning may be biased by the telescope observation axes acceptance variations. The central axes have a higher sensitivity than the border ones. This probably explains why we could not extract any fluctuations on the telescope scanning window borders (Fig.~\ref{figure_8})~: the associated axes have a weaker flux and thus a smaller weight in the PCA decomposition.

More constrains could be brought to the dynamics of the shallow hydrothermal system of the La Soufri\`ere of Guadeloupe's lava dome with additionnal cosmic muon telescopes deployed around the volcano to provide radiographies taken under different angles of view, and also using recent 3D structural imaging of the volcanic dome \cite{rosas-carbajal_volcano_2016}. This will allow to perform 3D reconstruction of the volumes $\mathcal{E}_1$, $\mathcal{E}_2$ and $\mathcal{E}_3$ and reduce the ranges of mass changes in Fig.~\ref{figure_6}. An even better reconstruction could be reached by merging the muon radiography measurements with continuous gravimetry data on the dome summit \cite{jourde_improvement_2015}. We estimated that the mass changes discussed in this study would generate gravity anomalies around a hundred microgal, which would be easy to measure with a standard gravimeter.

To conclude, we demonstrated that muon radiography provides unprecedented knowledge about the La Soufri\`ere of Guadeloupe shallow hydrothermal system dynamics. We hope that it will become a standard geophysical monitoring technique to both improve our understanding of the physical processes at work and bring useful informations to assess hazard level. For example it could become a efficient proxy to localize pressurized reservoirs and evaluate their mechanical stability in order to prevent potential volcanic blasts.


\section*{Methods -- Pixel merging through principal components analysis}

\begin{figure}
\begin{center}
\includegraphics[width=1.0\linewidth]{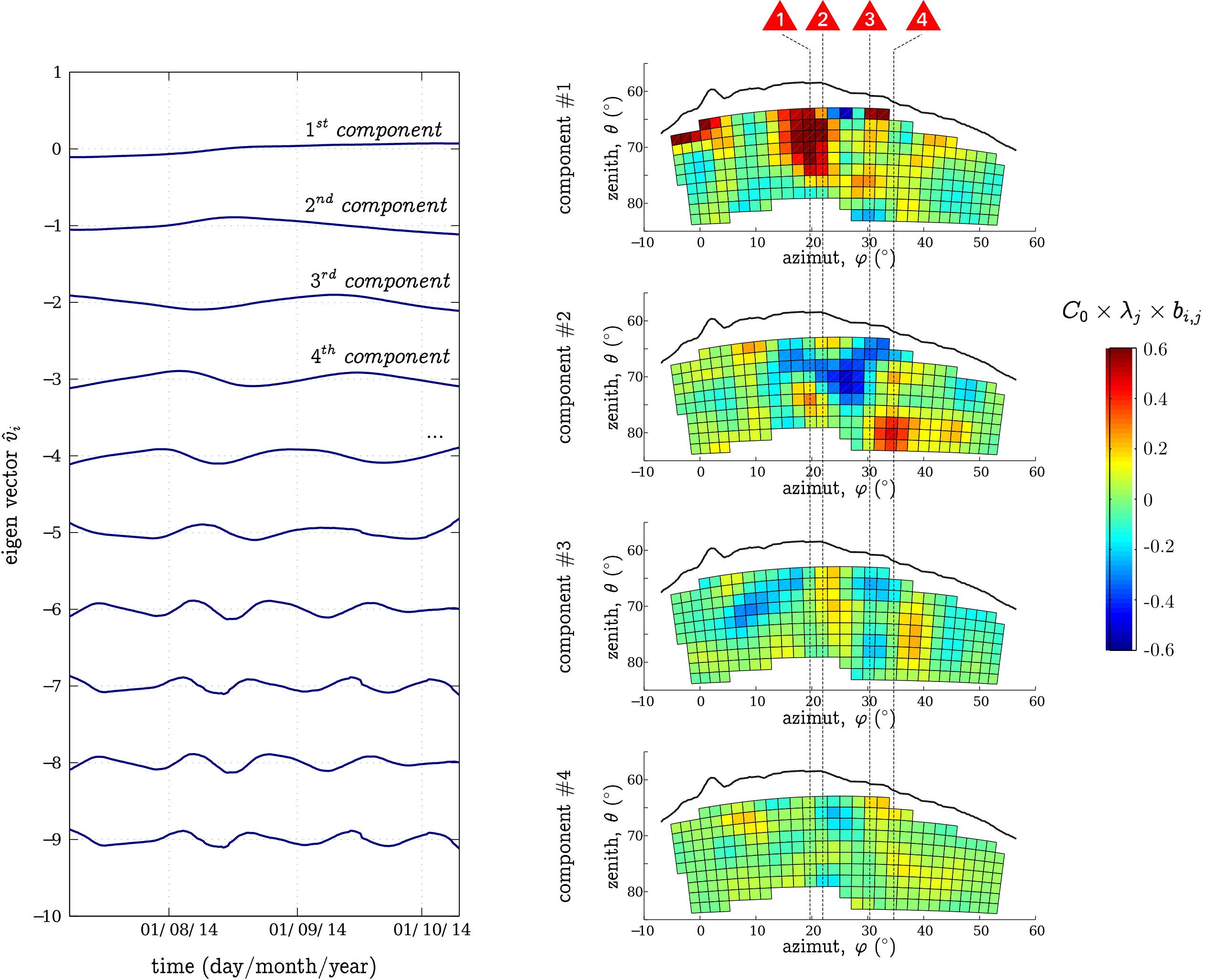}
\end{center}
\caption{\textit{Left~:} The first ten eigenvectors $\hat{v}_i$ extracted from the Summer 2014 muon tomography data. All the eigenvectors should be centred but here we added an increasing offset so that the trends do not superimpose on each other. To construct them each observation axis was previously smoothed using a 30 days large Hamming moving window. \textit{Right~:} Graphical representation of the four first eigenvector contributions to the measured signal. Each box corresponds to an independent observation axis $i$ and the associated color relates to $\lambda_j \times b_{i,j}$ ($j$ going from 1 to 4). $C_0$ is a scaling constant. The black solid curve delimits the interface between the volcanic dome and the sky. The red triangles refer to the different active zones visible at the surface of the dome azimuths (Fig.~\ref{figure_1}).}
\label{figure_8}
\end{figure} 

As explained previously, a telescope single observation axis does not collect enough particles to statistically distinguish variations in the muon flux linked to the volcanic activity on a monthly time-scale. We can solve the problem regrouping various observation axes in order to increase the signal intensity. Thus, we come up with the following two-folds problematic~: how to regroup the observation axes in order to efficiently extract temporal trends without introducing any prior information that may bias the solution~?

We propose here a solution using Principal Components Analysis (PCA, also called the Karhunen \& Loeve Transformation \cite{pearson_lines_1901, lebart_statistique_2006}). Let us note $\mathbf{s} (t) = (s_1 (t) , s_2 (t) , ... , s_N (t) )$, where $\mathbf{s} (t)$ is a vector containing the Summer 2014 temporal signals $s_i (t)$ associated to the $N$ different observation axes that cross the volcano. PCA gives an optimal linear decomposition of the $s_i (t)$ on an orthonormal basis $\hat{\mathbf{s}} (t) = (\hat{s}_1 (t) , \hat{s}_2 (t) , ... , \hat{s}_N (t) )$ calculated iteratively from the signal itself,

\begin{itemize}
\item \textbf{Iteration 1~:} we search $\hat{s}_1 (t)$ such as,
\begin{equation}
\hat{s}_1 (t) = \sum_{j = 1}^{N} a_{1,j} \times s_j (t),
\label{pca_eq_1}
\end{equation}
where the coefficients $a_{1,j}$ are estimated by minimizing the quadratic norm $\epsilon_1$,
\begin{equation}
\epsilon_1^2 = \sum_{j = 1}^{N} \parallel \hat{s}_1 (t) - s_j (t) \parallel^2.
\label{pca_eq_2}
\end{equation}
\item \textbf{Iteration i~:} we search $\hat{s}_i (t) ~ (1 < i \leqslant N)$ such as,
\begin{equation}
\hat{s}_i (t) = \sum_{j = 1}^{N} a_{i,j} \times s_j (t),
\label{pca_eq_3}
\end{equation}
where the coefficients $a_{i,j}$ are estimated by minimizing the quadratic norm $\epsilon_i$,
\begin{equation}
\epsilon_i^2 = \sum_{j = 1}^{N} \parallel \hat{s}_i (t) - ( \sum_{k = 1}^{i-1} s_j (t) - < s_j (t) , \hat{s}_k (t) > ) \parallel^2.
\label{pca_eq_4}
\end{equation}
The resulting basis is orthogonal, we normalize it,
\begin{equation}
\hat{s}_i (t) = \lambda_i \times \hat{v}_i (t) ~ \mathrm{, ~~ with ~} < \hat{v}_i , \hat{v}_j > = \delta_{i,j},
\label{pca_eq_5}
\end{equation}
where $\delta_{i,j}$ is the Kronecker Delta. $\hat{v}_i$ is the $i^{th}$ eigenvector and $\lambda_i$ its associated eigenvalue.
\end{itemize}

Finally, usinq eq.~\ref{pca_eq_3} and \ref{pca_eq_5} we get the decomposition,
\begin{equation}
s_i (t) = \sum_{j = 1}^{N} \left( \lambda_j \times b_{i,j} \right) \times \hat{v}_j (t),
\label{pca_eq_6}
\end{equation}
where $b_{i,j}$ are the square matrix $\mathbf{B}$ coefficients, with $\mathbf{B} = \mathbf{A}^{-1}$, $\mathbf{A}$ being the $N \times N$ square matrix associated to the $a_{i,j}$ coefficients.

Let us recall that PCA, as shown in eq.~\ref{pca_eq_4}, assumes the signals $s_i (t)$ are affected by a Gaussian noise. This condition is not fulfilled for all the telescope observation axes going through the volcano, and for the time-scale we are interested in, because of the muon flux weak intensity. In order to overcome this problem, we merge observation axes into groups of four neighbours and the flux is computed with a 30 days large Hamming moving window.

By construction, the eigenvectors are less and less representative of the input data ($\lambda_i$ is decreasing with i). Usually the first ones represent the global trends that are redundant on the different input signals while the last ones permit to reconstruct the little and specific discontinuities (mainly the noise). The eigenvectors are interesting four our study, because there is a possibility that they characterize different physical processes occurring inside the volcano, but we can also use the coefficients $b_{i,j}$ to map their respective contribution in the different regions of the dome.

Fig.~\ref{figure_8} shows the first ten eigenvectors (on the left) and, for the first four, their associated contribution on the different observation axes as seen from the telescope (on the right). Only the two first eigenvectors show fluctuations occurring on time-scales larger than a month, which can statistically be recovered considering the muon flux intensity (see the statistical time resolution section). The next eigenvectors present quicker fluctuations that characterize the statistical noise. Very clear large regions appear on the two first contribution maps (associated with the two first eigenvectors). The first component reveals a large coherent domain $\mathcal{E}_1$ aligned with the Tarissan pit (TP) and the North-Napol\'eon fume (NN). The second component shows two independent and anti-correlated regions~: a V-shape zone $\mathcal{E}_2$ which left arm is aligned with TP and NN, and right arm with the Crat\`ere Sud, the Gouffre Breislack (GB) and the Gouffre 56 (G56), and a bone-like zone $\mathcal{E}_3$ aligned with GB and G56. The next maps do not present any clear zone, the coefficients are small and appear to be randomly distributed in between the different observation axes (the little spatial coherence that appears is due to the neighbours merging mentioned previously).





\section*{Acknowledgements}
This study is part of the DIAPHANE project ANR-14-CE 04-0001. We acknowledge the financial support from the UnivEarthS Labex program of Sorbonne Paris Cit\'e (\textsc{anr-10-labx-0023} and \textsc{anr-11-idex-0005-02}). This is IPGP contribution ****.


\section*{Author contributions statement}
KJ, JM and DG conceived the experiment; KJ, JM, DG and JDB designed and constructed the apparatus; KJ, DG and JM conducted the experiment; KJ, JM, DG analysed the data; all authors wrote the article.


\section*{Additional information}
\textbf{Competing financial interests:} The authors declare no competing financial interests.

\end{document}